\documentclass[]{aastex631}

\shorttitle{Reltivistic jets from slim disks}
\shortauthors{Inoue}

\graphicspath{{./}{figures/}}

\begin{document}

\title{Relativistic jet ejections from slim disks}

\author{Hajime Inoue}
\affiliation{Institute of Space and Astronautical Science, Japan Aerospace Exploration Agency \\ 3-1-1 Yoshinodai, Chuo-ku, Sagamihara \\ Kanagawa 252-5210, Japan}

\begin{abstract}
An ejection mechanism of relativistic jets from slim disks is studied.
Since the radiation pressure is dominant in the slim disk, radiative energy flow arises along the pressure gradient in the vertical direction.
The divergence of the radiative flux tells us that the radiative energy flow from a bottom layer near the equatorial plane is absorbed by another layer upper than the boundary surface.
The absorbed energy accumulates in the upper layer as the matter advances inward and calculations show that the specific energy of the flow in the upper layer can be as large as $\sim c^{2}$ near the black hole when the accretion rate through the upper layer is relatively low.
Since the specific energy $\sim c^{2}$ is much larger than the gravitational energy, the height of the upper layer could significantly increase then.
Hence, the innermost part of the upper layer after almost all the angular momentum has been removed could have a much larger height than the black hole size and collide with one another around the central axis of the disk, bouncing back from the axis as simultaneously expanding along the axis.
The flow is expected to go outward along the central axis and to become supersonic due to the cross-section-change of the flow, getting the relativistic jets finally.

\end{abstract}


\section{Introduction}\label{Introduction}
In early 1990s, some X-ray persistent sources were discovered to exhibit radio images consisting of compact components at the center and two-sided jets, and, furthermore, superluminal motions were  detected from some recurrent transient X-ray sources (see Mirabel \& Rodriguez 1999 for the review of relativistic jet sources in the Galaxy).  

These discoveries stimulated observational studies of relativistic jets in transient X-ray sources, and, based on them,
Fender, Belloni \& Gallo (2004) presented a unified semiquantitative model for the disk - jet coupling in black hole binary systems.
According to their studies, it is suggested that there could exist two types of jets: steady jets and transient jets.
We recently discussed scenarios for the steady jets (Inoue 2022c).
A scenario for the transient jets was also proposed (Inoue 2023) but the paper was retracted because errors were found in the considerations (PASJ, 75, 833, 2023). 
We revisit the transient jet phenomenon and present a new scenario for it here.

One of the basic issues of the jet ejections is that the specific energy of the ejected matter is required to be positive to get to infinity, while that of the accreted matter is naively considered to be negative at the outermost boundary of the accretion flow.
Although the energy is possible to be supplied by the black hole (e.g. Blandford \& Znajek 1977), the coupling of the jet ejections with the accretion disk states observed from the black hole binaries strongly suggests that the energy source would be the accretion flow itself.
Narayan and Yi (1994) pointed out that the Bernoulli parameter representing the sum of the specific kinetic, thermal and gravitational energies of the accreted matter can be positive in the typical region of the advection dominated accretion flow (ADAF) even if the total specific energy is negative taking into account the outward energy flow associated with the angular momentum transfer.
Utilizing the possibility for the accreted matter to escape from the gravitational potential under the positive Bernoulli parameter, Blandford and Begelman (1999) proposed a model, called as advection-dominated inflow-outflow solutions (ADIOS).
In parallel, Ferreira (1997) studied magnetically-driven jets and a series of papers on a unified framework of the jet-emitting disk - standard accretion disk (JED-SAD) paradigm has followed (see Marcel et al. 2022 and references therein).
On the other hand, behaviors of the transonic flow from the outer subsonic region to the inner supersonic 
region in an ADAF have been studied in many papers (see e.g. chapter 8 in Kato et al. 2008; a brief review in the introduction section of Kumar \& Chattopadhyay 2013).
An interesting argument among them is that a shock can appear between two critical points from a subsonic flow to a supersonic flow and a mass outflow can be induced behind the shock (see Kumar \& Chattopadhyay 2013; 2017 and references therein). 
In addition, Inoue (2022c) recently proposed a situation for the jet matter to get the positive specific energy where two layers exist in the accretion flow and one layer receives energy from the other such that the specific energy becomes positive.
As briefly reviewed above, a number of studies have been done for the jet ejection mechanism.
However, we are still far from complete resolution of the ejection mechanism of the transient relativistic jets.


In this paper, we propose an ejection-mechanism of relativistic jets from slim disks.
The basic idea is already proposed for the steady jet ejection from the slim disk by Inoue (2022c).
He considers a situation in which two (bottom and upper) layers exist in the slim disk and the upper layer receives radiative energy from the bottom layer such that the specific energy becomes positive to flow out to infinity. 
We apply the idea here to the transient relativistic jet ejections, studying the radiative energy transfer from the bottom layer to the upper layer more precisely.

Then, we discuss consistencies of predictions from the model with implications from  observations of the black hole binary GRS 1915+105 which has often exhibited transient jet ejections since the detection of the first galactic superluminal jets (Mirabel \& Rodriguez 1994).

Three types of accretion flow, the standard disk (Shakura \& Sunyaev 1973), the slim disk  (Abramovicz et al. 1988) and the hot accretion flow  (HAF) (e.g. Yuan \& Narayan 2014) are introduced to discuss configurations of the accretion flow in various states hereafter.
Observationally. the flow types are distinguished basically through the properties of the X-ray continuum spectrum which is usually approximated by a combination of the disk blackbody component (Mitsuda et al. 1984) and the power law component. 

Both the standard disk and the slim disk radiate the blackbody emission from the surface and can be responsible for the disk blackbody component.
The further observational distinction between the two is done with overall behaviors of X-ray intensity, spectrum, and their time variations (see e.g. Inoue 2022a).
An example of such a distinction is presented in Appendix \ref{PhaseIII}, which gives us the important observational indication to the present study.

The power law spectrum extending up to $\sim$100 keV requires a HAF containing such high energy electrons and being optically thin or gray, as the origin.
Different names can be used for the HAF depending on the situation, say ADAF (the advection-dominated accretion flow: Narayan and Yi  1994); LHAF (the luminous hot accretion flow : Yuan 2001); ADIOS (Blandford and Begelman 1999); JED (Ferreira 1997; Marcel et al. 2022 and references therein).
The present discussions are not influenced by the differences, and thus the term, HAF, is used as the phenomenological term for the accretion flow responsible for the power law component.


In this study, the central black hole is assumed to be the Schwarzschild one and $R_{\rm s}$ represents the Schwarzschild radius.
The velocity of light is expressed with $c$.
We apply the cylindrical coordinate ($r$, $\phi$, $z$) where $r$ is the radial distance along the equatorial plane and the $z$ axis is perpendicular to the equatorial plane.


\section{Scenario for transient jet ejection}\label{Scenario}
\subsection{Two stratified layers in the slim disk}
Properties of the slim disk are usually obtained with simplification due to integration along the $z$ direction (e.g. Abramovitz et al. 1988).
As for the energy equation, however, we start from the basic equation given as
\begin{equation}
\frac{\partial (r \rho v \varepsilon)}{r \partial r} = - \dot{U},
\label{eqn:EnergyFlowEq}
\end{equation}
where $\rho$ is the density, $v$ is the inflow velocity, and $\varepsilon$ is the specific energy of the inflowing matter.
$\dot{U}$ is the gain rate of the radiation energy density and can easily be calculated in the case of the slim disk as follows.

Since the radiation pressure is considered to be much stronger than the gas pressure in the slim disk, 
the mechanical balance in the $z$ direction can be approximated as
\begin{equation}
\frac{\partial P_{\rm rad}}{\partial z} = -\rho \frac{GMz}{(r^{2}+z^{2})^{3/2}},
\label{eqn:dPrdz}
\end{equation}
where $P_{\rm rad}$ is the radiation pressure, $G$ is the gravitational constant and $M$ is the mass of the central black hole.
On the other hand, the radiation flux in the $z$ direction, $F_{\rm z}$, is given as
\begin{equation}
F_{\rm z} =-\frac{c}{3\kappa_{\rm T} \rho} \frac{\partial U}{\partial z},
\label{eqn:PhotonDiffEq}
\end{equation}
where $U$ is the radiation energy density and $\kappa_{\rm T}$ is the opacity for the Thomson scattering.
Considering the relation of $P_{\rm rad} = U/3$, we obtain from above two equations 
\begin{equation}
F_{\rm z} = \frac{cGM}{\kappa_{\rm T}} \frac{z}{(r^{2}+z^{2})^{3/2}}.
\label{eqn:F_z}
\end{equation}
The radiative energy gain rate per volume, $\dot{U}$, can be calculated from the divergence of the radiative flux, $F_{\rm z}$, as
\begin{equation}
\dot{U} = - \frac{\partial F_{\rm z}}{\partial z} = -\frac{cGM}{\kappa_{\rm T}} \frac{r^{2} - 2z^{2}}{(r^{2}+z^{2})^{5/2}}.
\label{eqn:dotU}
\end{equation}

We see from this equation that $\dot{U} < 0$ in the region of $z < z_{\rm b}$, while $\dot{U} > 0$ in the region of $z > z_{\rm b}$, where 
\begin{equation}
z_{\rm b} = r/\sqrt{2}.
\label{eqn:z_b}
\end{equation}
As also seen from Equation (\ref{eqn:dotU}),
$F_{\rm z}$ increases as $z$ increases in the bottom region below $z_{\rm b}$ and the flux increment should need energy supply from the radiation field on the way of the $F_{\rm z}$ increase.
On the other hand, $F_{\rm z}$ decreases as $z$ increases in the upper region above $z_{\rm b}$ and the flux decrement should be due to energy absorption by the radiation field on the way of $F_{\rm z}$ decrease.
As the result, an amount of energy is radiatively transfered from the bottom region to the upper region. 
This is simply due to the particular situation of the radiation-pressure dominated and the Thomson-scattering dominated atmosphere in which the radiative diffusion rate is directly determined by the strength of the gravitational acceleration as seen from Equation (\ref{eqn:F_z}).
Since the gravitational acceleration at the top surface of the bottom region is larger than that of the upper region, the radiative flux from the upper region becomes smaller than that from the bottom region and the decrement should be the result of the absorption by the upper region.


If the matter in the slim disk flows along a hollow cone surface with a constant $z/r$, the flow can be separated into two layers, where matter in one layer with $z < z_{\rm b}$ continuously loses the energy and that in the other layer with $z > z_{\rm b}$ continuously gains the energy as the flow advances inward.  
Hereafter, we call the energy losing and gaining layers as the bottom layer and the upper layer respectively.

\subsection{Specific energy of matter in the innermost region of slim disk}
Now, let us consider the average properties of the energy flow in the $-r$ direction of each of the two layers.

\begin{figure}
 \begin{center} 
  \includegraphics[width=10cm]{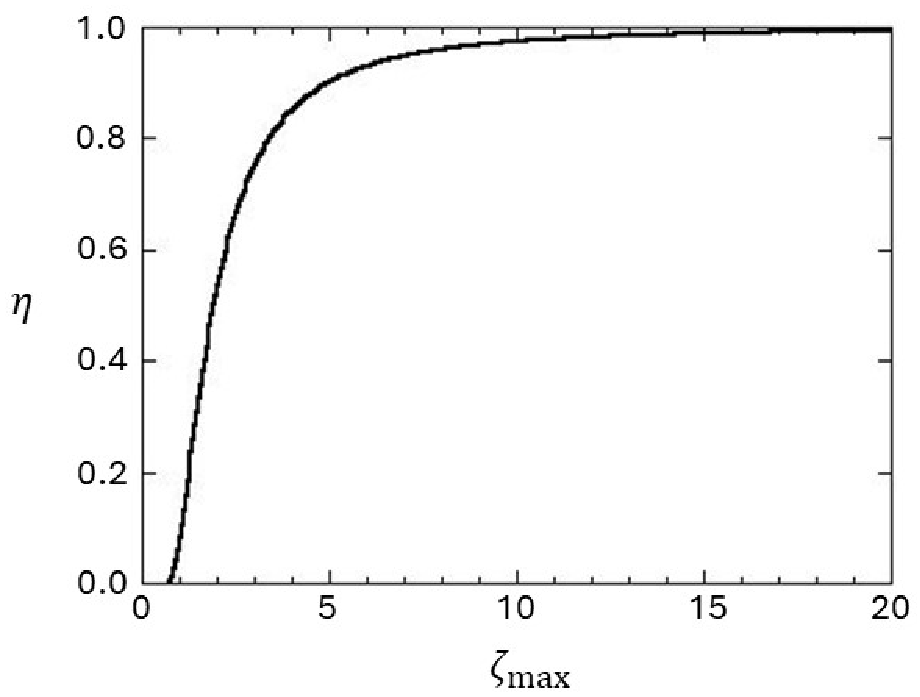} 
 \end{center}
\caption{Fraction of the energy absorption rate by the upper layer to the radiative energy flux from the bottom layer, $\eta$, versus the maximum height of the upper layer normalized by $r$, $\zeta_{\rm max}$.}\label{fig:AbsEff}
\end{figure}

Integrating Equation (\ref{eqn:EnergyFlowEq}) over $z$, with the help of Equation (\ref{eqn:dotU}), from 0 to $z_{\rm b}$ for the bottom layer and from $z_{\rm b}$ to the maximum $z$ position, $z_{\rm max}$, for the upper layer, we obtain the following two equations as
\begin{equation}
\frac{d (r \int_{0}^{z_{\rm b}} \rho v \varepsilon dz)}{r d r} = \dot{W}_{btm}^{(+)},
\label{eqn:EnergyFlowEq-btm}
\end{equation}
for the bottom layer and
\begin{equation}
\frac{d (r \int_{z_{\rm b}}^{z_{\rm max}} \rho v \varepsilon dz)}{r d r} = - \dot{W}_{\rm up}^{(-)},
\label{eqn:EnergyFlowEq-up}
\end{equation}
for the upper layer.
Here, $\dot{W}_{\rm btm}^{(+)}$ is the radiative flux from the bottom layer, and is calculated as 
\begin{equation}
\dot{W}_{\rm btm}^{(+)} = \int_{0}^{z_{\rm b}} \dot{U} dz = F_{\rm z}(z_{\rm b}) = 
\frac{2}{3^{3/2}}\frac{cGM}{\kappa_{\rm T}} \frac{1}{r^{2}}.
\label{eqn:Wdot-btm}
\end{equation}
On the other hand, $\dot{W}_{\rm up}^{(-)}$ is the energy absorption rate by a column with unit cross section of the upper layer, and can be written as
\begin{equation}
\dot{W}_{\rm up}^{(-)} = - \int_{z_{\rm b}}^{z_{\rm max}} \dot{U} dz = F_{\rm z}(z_{\rm b}) - F_{\rm z}(z_{\rm max}) = \eta \dot{W}_{\rm btm}^{(+)},
\label{eqn:Wdot-up}
\end{equation}
where
\begin{equation}
\eta = 1 - \frac{3^{3/2}}{2}  \frac{\zeta_{\rm max}}{(1+\zeta_{\rm max}^{2})^{3/2}} \ \ \mbox{ for } \zeta_{\rm max} \ge 1/\sqrt{2},
\label{eqn:eta}
\end{equation}
and $\zeta_{\rm max} \equiv z_{\rm max}/r$.
The parameter $\eta$ expresses how much fraction of the radiative energy flux from the bottom layer is absorbed by the upper layer, and $(1 - \eta) \dot{W}_{\rm btm}^{(+)} = F(z_{\rm max})$ is the radiative flux from the top surface of the upper layer.
The value of $\eta$ is plotted against $\zeta_{\rm max}$ in Figure \ref{fig:AbsEff}.

The maximum height $z_{\rm max}$ determines the value of $\eta$.
If we employ an approximation of the adiabatic relation, $P \propto \rho^{\gamma}$, the structure of the slim disk given by Equation (\ref{eqn:dPrdz}) can be solved.
The solution for the density at a radial distance $r$ is described as a function of $z$ as
\begin{equation}
\frac{\rho}{\rho_{0}} = \left[ 1 - \frac{(\gamma - 1) GM}{c_{\rm s, 0}^{2} r} \left\{ 1 - \frac{1}{(1 + \zeta^{2})^{1/2}}\right\} \right]^{1/(\gamma-1)},
\label{eqn:rho}
\end{equation}
where $\rho_{0}$ and $c_{\rm s, 0}$ are the density and the sound velocity at $z = 0$.
On the other hand, the equation of specific energy conservation for the slim disk approximately gives us a relation in case of $r \gg R_{\rm s}$ as
\begin{equation}
\frac{1}{\gamma-1}c_{\rm s, 0}^{2} \simeq \frac{3}{2} \frac{GM}{r},
\label{eqn:c_s0}
\end{equation}
taking account of the outward energy flow associated with the angular momentum transfer and assuming the  Keplerian circular rotation of the disk matter.
By substituting Equation (\ref {eqn:c_s0}) into Equation (\ref{eqn:rho}), we see that the density of the slim disk extends to infinity in the $z$ direction mathematically.
In practice, however, the maximum height, $z_{\rm max}$ could be limited by a value as the sound velocity times the matter in-falling time, $t_{\rm if}$.
Since $t_{\rm if} \sim r/v$ with the in-falling velocity, $v$, which is roughly given  from the standard disk theory to be $\alpha c_{\rm s}^{2}/v_{\phi}$ in terms of the viscous parameter $\alpha$ (Shakura \& Sunyaev 1973) and the rotational velocity $v_{\phi}$, we can obtain an approximate equation, with help of Equation (\ref{eqn:c_s0}) and $\gamma = 4/3$, as
$\zeta_{\rm max} \sim \sqrt{2}/\alpha$, which is calculated to be $\sim 14$ and $5$ for $\alpha = 0.1$ and 0.3 respectively. 
$\eta$ is very close to 1 for $\zeta_{\rm max} = 14$, and is $\sim 0.9$ even for   $\zeta_{\rm max} = 5$, as seen from Figure \ref{fig:AbsEff}.
As discussed later, it is expected in the present study that the sound velocity in the upper layer gets larger and thus the maximum disk height becomes higher than those estimated above.
Hence, we assume $\eta = 1$, namely that all the radiative flux from the bottom layer is absorbed by the upper layer, hereafter.

Next, let us multiply the both sides of the above two equations by 4$\pi$, integrating them over $r$ from the 
minimum radius, $r_{\rm min}$,
to the maximum radius, $r_{\rm max}$, 
of the slim disk, and approximate
\begin{equation}
4\pi r \int_{0}^{z_{\rm b}} \rho v \varepsilon dz \simeq \left( 4\pi r \int_{0}^{z_{\rm b}} \rho v dz \right) \bar{\varepsilon}_{\rm btm} = \dot{M}_{\rm btm} \bar{\varepsilon}_{\rm btm}
\label{eqn:MdotEpsilon_btm}
\end{equation}
and 
\begin{equation}
4\pi r \int_{z_{\rm b}}^{z_{\rm max}} \rho v \varepsilon dz \simeq \left( 4\pi r \int_{z_{\rm b}}^{z_{\rm max}} \rho v dz \right) \bar{\varepsilon}_{\rm up} = \dot{M}_{\rm up} \bar{\varepsilon}_{\rm up},
\label{eqn:MdotEpsilon_btm}
\end{equation}
where $\dot{M}$ and $\bar{\varepsilon}$ respectively represent the accretion rate and the average specific energy over the relevant layer.
Then, we get
\begin{equation}
\dot{M}_{\rm btm, out} \bar{\varepsilon}_{\rm btm, out} - \dot{M}_{\rm btm, in} \bar{\varepsilon}_{\rm btm, in} \simeq \frac{2}{3^{3/2}} \ln \left(\frac{r_{\rm max}}{r_{\rm min}}\right)  L_{\rm E},
\label{eqn:EnergyTransfer-btm}
\end{equation}
and
\begin{equation}
\dot{M}_{\rm up, out} \bar{\varepsilon}_{\rm up, out} - \dot{M}_{\rm up, in} \bar{\varepsilon}_{\rm up, in} \simeq - \frac{2}{3^{3/2}} \ln \left(\frac{r_{\rm max}}{r_{\rm min}}\right) L_{\rm E},
\label{eqn:EnergyTransfer-btm}
\end{equation}
where $L_{\rm E} \equiv 4\pi cGM/\kappa_{\rm T}$ is the Eddington luminosity. The subscripts, btm, up, in and out indicate quantities of the bottom layer, the upper layer, the innermost region and the uppermost region, respectively.

Finally, approximating $\bar{\varepsilon}_{\rm btm, out} = \bar{\varepsilon}_{\rm up, out} = 0$, the specific energies of the matter in the innermost regions of the two layers are represented as
\begin{equation}
\bar{\varepsilon}_{\rm btm, in} \simeq - \frac{2}{3^{3/2}} \ln \left(\frac{r_{\rm max}}{r_{\rm min}}\right) \frac{L_{\rm E}}{\dot{M}_{\rm btm, in}} = - \frac{\dot{M}_{\rm c}}{\dot{M}_{\rm btm, in}} c^{2},
\label{eqn:varepsilon-btm}
\end{equation}
and
\begin{equation}
\bar{\varepsilon}_{\rm up, in} \simeq \frac{2}{3^{3/2}} \ln \left(\frac{r_{\rm max}}{r_{\rm min}}\right) \frac{L_{\rm E}}{\dot{M}_{\rm up, in}} =  \frac{\dot{M}_{\rm c}}{\dot{M}_{\rm up, in}} c^{2},
\label{eqn:varepsilon-up}
\end{equation}
where $\dot{M}_{\rm c}$ is defined as
\begin{equation}
\dot{M}_{\rm c} \equiv \frac{2}{3^{3/2}} \ln \left(\frac{r_{\rm max}}{r_{\rm min}}\right) \frac{L_{\rm E}}{c^{2}},
\label{eqn:M_c}
\end{equation}
and is calculated to be $1.2 \sim 2.5 \times 10^{18}$ g s$^{-1}$ for $M = 10\ M_{\odot}$ and a range of $r_{\rm max}/r_{\rm min}$ from 10 to $10^{2}$.

Here, $r_{\rm min}$ is considered to correspond to the distance of the sonic point inner than which the slim disk becomes supersonic flow toward the black hole, and be $\sim 2 R_{\rm s}$ (see e.g. Inoue 2022c).
On the other hand, $r_{\rm max}$ could be regarded as the boundary position inside of which the gas-pressure-dominated standard disk changes to the radiation-pressure-dominated standard disk which is unstable so that the disk immediately turns to the slim disk.
According to the standard disk model, the boundary distance, $r_{\rm b}$, is given as
\begin{equation}
r_{\rm b} \simeq 18 \left(\frac{\alpha}{0.1}\right)^{2/21} \left(\frac{M}{10\ M_{\odot}}\right)^{2/21} \left(\frac{\dot{M}}{\dot{M}_{\rm E}}\right)^{16/21}\ R_{\rm s}
\label{eqn:r_b}
\end{equation}
 (see p.149 in Kato et al. 2008), where $\dot{M}_{\rm E}$ is defined to be $L_{\rm E}/c^{2}$ and is $\sim 1.4 \times 10^{18}$ g s$^{-1}$ for $M = 10 M_{\odot}$.
Observationally, the slim disk appears when the accretion rate toward a black hole  exceeds $\dot{M}_{\rm E}$ as in the very high state (see e.g. Inoue 2022a).
Thus, we can see from the above equation that $r_{\rm max} \sim r_{\rm b} \gtrsim 20 R_{\rm s}$ and then $r_{\rm max}/r_{\rm min} \gtrsim 10$.
Since significant jet ejections are expected from the slim disk with relatively low accretion rate as discussed later, we have calculated the $\dot{M}_{\rm c}$ value for the smallest range of $r_{\rm max}/r_{\rm min}$ from 10 to 10$^{2}$ above.

As seen from Equation (\ref{eqn:varepsilon-up}), the specific energy of the upper layer can be as large as $\sim c^{2}$ or larger when $\dot{M}_{\rm up, in}$ is around $\dot{M}_{\rm c}$ or less. 

\subsection{Relation of the accretion rates between the upper layer and the bottom layer}\label{AccretionRateRelation}
Now, we approximately see the density distributions of the two layers.

The solution for the density in the bottom layer in the innermost region with the typical distance, $r_{\rm in}$, is  described by Equation (\ref{eqn:rho}) after replacing $\rho$ and $r$ with $\rho_{\rm btm}$ and $r_{\rm in}$ respectively.
Then, we can relate $c_{\rm s, 0}$ to the average specific thermal energy (enthalpy) of the bottom layer, $\bar{\varepsilon}_{\rm th, btm, in}$, by an approximation as
\begin{equation}
c_{\rm s, 0}^{2} \simeq (\gamma-1) \bar{\varepsilon}_{\rm th, btm, in},
\label{eqn:c_s_0-epsilon}
\end{equation}
and obtain $\bar{\varepsilon}_{\rm th, btm, in}$ with the help of Equation (\ref{eqn:varepsilon-btm}) as a function of $\dot{M}_{\rm btm, in}$ as
\begin{equation}
\bar{\varepsilon}_{\rm th, btm, in} \simeq \bar{\varepsilon}_{\rm th, 0} - \bar{\varepsilon}_{\rm btm, in} \simeq \left( \frac{R_{\rm s}}{4 r_{\rm in}} - \frac{\dot{M}_{\rm c}}{\dot{M}_{\rm btm, in}}\right) c^{2}.
\label{eqn:Epsilon-Mdot-btm}
\end{equation}
Here, $\bar{\varepsilon}_{\rm th, 0}$ is  the specific thermal energy of a slim disk around $r_{\rm in}$ in case without energy transfer, which can be approximately given as
\begin{equation}
\bar{\varepsilon}_{\rm th, 0} \simeq \frac{GM}{2r_{\rm in}} = \frac{R_{\rm s}}{4 r_{\rm in}} c^{2}.
\label{eqn:Epsilon_th_0}
\end{equation}

For a given $\dot{M}_{\rm btm, in}/\dot{M}_{\rm c}$, we can get the value of $c_{\rm s, 0}^{2}$ from Equations (\ref{eqn:c_s_0-epsilon}) and (\ref{eqn:Epsilon-Mdot-btm}), and then calculate 
a non-dimensional parameter, $\psi_{\rm btm}$, defined as
\begin{equation}
\psi_{\rm btm} \equiv \int_{0}^{\zeta_{\rm b}} \frac{\rho_{\rm btm}}{\rho_{0}} d\zeta,
\label{eqn:psi-Def}
\end{equation}
where $\zeta_{\rm b} \equiv z_{\rm b}/r_{\rm in}$.  $\rho_{\rm btm}/\rho_{0}$ is given in Equation (\ref{eqn:rho}) for $\rho = \rho_{\rm btm}$ and $r = r_{\rm in}$. 
This parameter is the non-dimensional representative of the half column density of the bottom layer, which is $\rho_{0} r_{\rm in} \psi_{\rm btm}$.

In terms of this half column density, the accretion rate of the bottom layer is described as 
$\dot{M}_{\rm btm, in} \simeq 4 \pi r_{\rm in}^{2} v \rho_{0} \psi_{\rm btm} $, where $v$ is the in-falling velocity of the bottom layer.
Then, if we express $\dot{M}_{\rm c}$ as $\dot{M}_{\rm c} = 4 \pi r_{\rm in}^{2} v \rho_{\rm ref}$, introducing a reference density, $\rho_{\rm ref}$, 
$\dot{M}_{\rm btm,in}/\dot{M}_{\rm c}$ is represented as
\begin{equation}
\frac{\dot{M}_{\rm btm, in}}{\dot{M}_{\rm c}} = \frac{\rho_{0}}{\rho_{\rm ref}} \psi_{\rm btm}.
\label{eqn:dotM_btm/dotM_c-psi}
\end{equation}
From this equation, we can obtain the normalized value of $\rho_{0}$ from the given $\dot{M}_{\rm btm, in}/\dot{M}_{\rm c}$ and the calculated value of $\psi_{\rm btm}$.

\begin{figure}
 \begin{center} 
  \includegraphics[width=12cm]{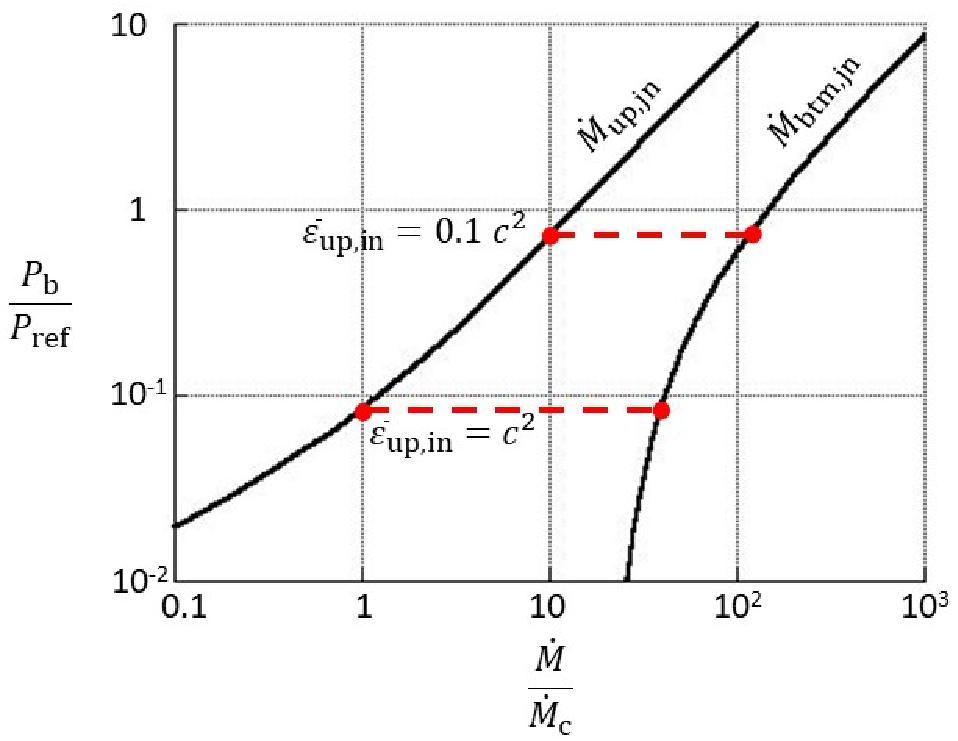} 
 \end{center}
\caption{The normalized pressures of the upper layer and the bottom layer at the boundary as a function of $\dot{M}/\dot{M}_{\rm c}$.  The specific energy of the upper layer, $\varepsilon_{\rm up, in}$, is estimated to be $\dot{M}_{\rm c}/\dot{M}_{\rm up, in}$ times $c^{2}$.  Two cases of the combinations of the accretion rates of the two layers are shown: one for $\bar{\varepsilon}_{\rm up, in} = c^{2}$ and the other for $\bar{\varepsilon}_{\rm up, in} = 0.1 c^{2}$. }\label{fig:Mdot-Pb}
\end{figure}

Now that the density distribution has been obtained for the given $\dot{M}_{\rm btm, in}/\dot{M}_{\rm c}$, we can calculate the pressure of the bottom layer at the boundary to the upper layer at $z_{\rm b}$, which is given as 
\begin{equation}
P_{\rm btm, b} = \frac{1}{\gamma} \rho_{0} c_{\rm s, 0}^{2} 
\left( \frac{\rho_{\rm btm, b}}{\rho_{0}}\right)^{\gamma},
\label{eqn:P_b-btm}
\end{equation}
where $\rho_{\rm btm,b}$ is the density of the bottom layer at $z_{\rm b}$ calculated from Equation (\ref{eqn:rho}) for $\rho = \rho_{\rm btm}$ and $r = r_{\rm in}$.
The pressure of the bottom layer at $z_{\rm b}$ normalized by the reference pressure, $P_{\rm ref} \equiv \rho_{\rm ref} c^{2}/\gamma$ in case of $r_{\rm in} = 3 R_{\rm s}$ and $\gamma = 4/3$ is plotted against $\dot{M}_{\rm btm, in}/\dot{M}_{\rm c}$ in Figure \ref{fig:Mdot-Pb}.
As seen from this figure, when $\dot{M}_{\rm btm, in} \lesssim 20 \dot{M}_{\rm c}$, no solution for $P_{\rm btm, z}$ exists and thus the two layer situation does not appear.  
This is because the specific thermal energy is too low for the height of the bottom layer to exceed $z_{\rm b}$ in this low accretion rate range.

On the other hand, the specific thermal energy of the upper layer can be obtained with the help of Equation (\ref{eqn:varepsilon-up}) as a function of $\dot{M}_{\rm up, in}$ as
\begin{equation}
\bar{\varepsilon}_{\rm th, up, in} \simeq \bar{\varepsilon}_{\rm th, 0} + \bar{\varepsilon}_{\rm up, in} \simeq \left( \frac{R_{\rm s}}{4 r_{\rm in}} + \frac{\dot{M}_{\rm c}}{\dot{M}_{\rm up, in}}\right) c^{2}.
\label{eqn:Epsilon-Mdot-up}
\end{equation}
In this case, the specific thermal energy increases as the accretion rate decreases.

The density distribution of the upper layer is given on the adiabatic approximation for the upper layer as
\begin{equation}
\frac{\rho_{\rm up}}{\rho_{\rm up, b}} = \left[ 1 - \frac{(\gamma - 1) GM}{c_{\rm s, up, b}^{2} r_{\rm in}} \left\{ \frac{1}{(1 + \zeta_{\rm b}^{2})^{1/2}} - \frac{1}{(1 + \zeta^{2})^{1/2}}\right\} \right]^{1/(\gamma-1)},
\label{eqn:rho-up}
\end{equation}
where $\rho_{\rm up, b}$ and $c_{\rm s, up, b}$ are the density and the sound velocity at $z = z_{\rm b}$.
Although this mathematical solution extends to infinity when $(\gamma-1) GM/c_{\rm s, up, b}^{2} r_{\rm in} < 1$, namely $c_{\rm s. up. b}^{2} > (\gamma-1) GM/r_{\rm in}$, the practical distribution should be limited in a range to which the matter can reach within the in-falling time, $t_{\rm if}$.
The uppermost boundary of the upper layer, $z_{\rm max}$, could roughly be estimated to be a value of the sound velocity times $t_{\rm if}$ and we assume $z_{\rm max}/r_{\rm in} = 10 c_{\rm s, up, b} /c$ here.
The density distribution in Equation (\ref{eqn:rho}) gives a finite $z_{\rm max}/r_{\rm in}$ value of $\sqrt{3}$ for $c_{\rm s, 0} = c/6 $ in case of $r_{\rm in} = 3 R_{\rm s}$ and $\gamma = 4/3$.  The above assumption on $z_{\rm max}/r_{\rm in}$ can smoothly connect with this  $z_{\rm max}/r_{\rm in}$ value at the lowest end of $c_{\rm s}$.

Similarly to the case of the bottom layer, we can calculate the $c_{\rm s, up, b}$ value from Equation (\ref{eqn:Epsilon-Mdot-up}) for a given $\dot{M}_{\rm up, in}/\dot{M}_{\rm c}$, and the value of $\psi_{\rm up}$ introduced as
\begin{equation}
\psi_{\rm up} = \int_{\zeta_{\rm b}}^{\zeta_{\rm max}} \frac{\rho_{\rm up}}{\rho_{\rm up, b}} d\zeta,
\label{eqn:psi-up}
\end{equation}
utilizing Equation (\ref{eqn:rho-up}).  Here, $\zeta_{\rm max} \equiv z_{\rm max}/r_{\rm in}$.
Then, we can get the value of $\rho_{\rm up, b}/\rho_{\rm ref}$ from
\begin{equation}
\frac{\dot{M}_{\rm up, in}}{\dot{M}_{\rm c}} = \frac{\rho_{\rm up, b}}{\rho_{\rm ref}} \psi_{\rm up}.
\label{eqn:dotM_up/dotM_c-psi}
\end{equation}
for the given $\dot{M}_{\rm up, in}/\dot{M}_{\rm c}$ and the calculated $\psi_{\rm up}$ value.
Finally, the pressure of the upper layer at the boundary to the bottom layer can be obtained as
\begin{equation}
P_{\rm up, b} = \frac{1}{\gamma} \rho_{\rm up, b} c_{\rm s, up, b}^{2}
\label{eqn:P_b-up}
\end{equation}
The pressure of the upper layer at $z_{\rm b}$ normalized by $P_{\rm ref}$ in case of $r_{\rm in} = 3 R_{\rm s}$ and $\gamma = 4/3$ is plotted against $\dot{M}_{\rm up, in}/\dot{M}_{\rm c}$ also in Figure \ref{fig:Mdot-Pb}.

If the upper layer rides on the bottom layer, the pressures of the two layers should balance with each other at the boundary.
Hence, the relation of the accretion rate between the two layers can be seen by a combination of the two $\dot{M}$ values along a horizontal, constant $P_{\rm b}$ line.
For examples, two cases are indicated in the figure.
One case is for $\dot{M}_{\rm up, in} = \dot{M}_{\rm c}$ corresponding to $\bar{\varepsilon}_{\rm up, in} = c^{2}$.  In this case, $\dot{M}_{\rm btm, in}$ is $\sim 40 \dot{M}_{\rm c}$.
The other is for $\dot{M}_{\rm up, in} = 10 \ \dot{M}_{\rm c}$ corresponding to $\bar{\varepsilon}_{\rm up, in} = 0.1\ c^{2}$.  In this case, $\dot{M}_{\rm btm, in}$ is $\sim 110 \dot{M}_{\rm c}$.

\subsection{Conversion to the jet flow}\label{FlowConversion}
Supposing that the matter in the upper layer gets the specific thermal energy as large as $\sim c^{2}$, the energy is well larger the gravitational energy, and thus the height of the upper layer could increase as it advances inward. 
A small part of the flow in which the stream lines are directed to the black hole could be captured by the black hole.  However, the other large part is expected to collide one another around the $z$ axis well outside the black hole and to bounce back to the $+r$ direction as simultaneously expanding in the $z$ direction, if the angular momentum is still efficiently transferred outward in the $r$ direction.

\begin{figure}
 \begin{center} 
  \includegraphics[width=8cm]{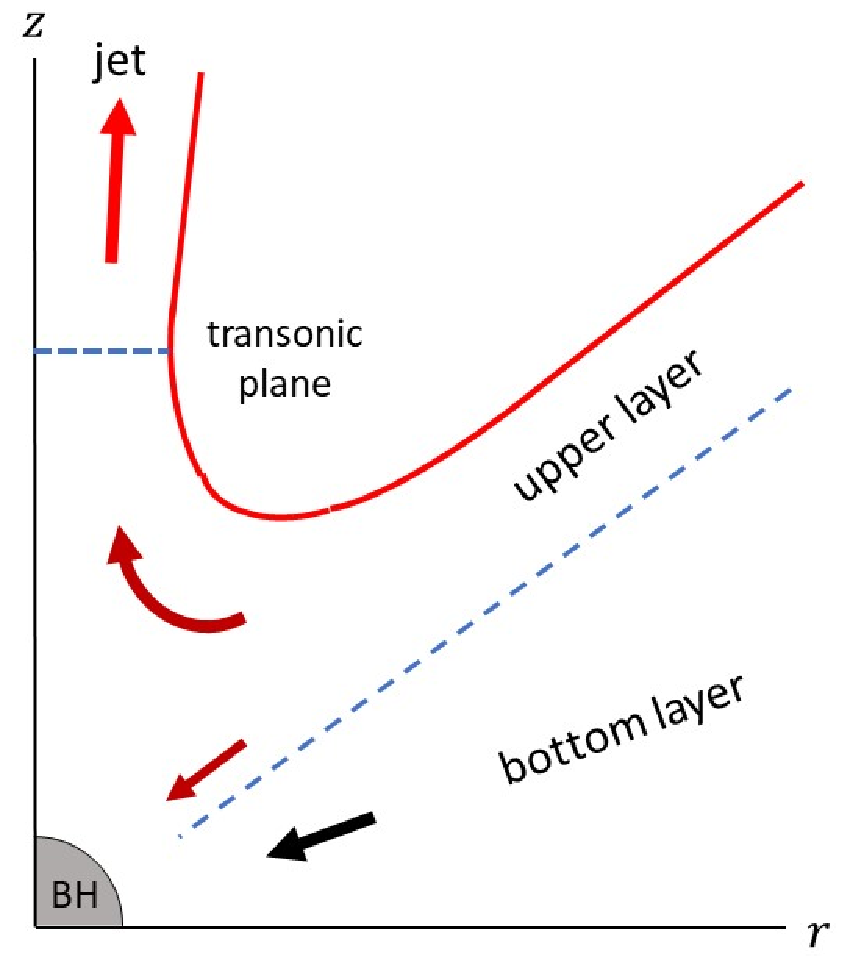} 
 \end{center}
\caption{Schematic cross section of the flow conversion from the slim disk to the jet.  The flow boundary is conjectured with the solid line.}\label{fig:JetScenario}
\end{figure}

As the result, we can expect that most of the flow in the upper layer turns to the outward flow along the $z$ direction.
Figure \ref{fig:JetScenario} shows the schematic cross section of the flow conversion of the flow in the upper layer to the jet flow.

As shown in appendix \ref{JetEjection}, the duration of the jet ejection is roughly a few sec.
Since this period is much longer than the dynamical time scale in which  the local dynamical balance in the flow is adjusted, the flow configuration as in Figure \ref{fig:JetScenario} can be well approximated to be in the steady state.

If we introduce  the area of the cross section of the outflow, $S$, as a function of $z$,  
$S$ is expected to decrease when $z$ is small but to come to increase after $z$ exceeds a certain position as seen in Figure \ref{fig:JetScenario}.
As indicated in Appendix \ref{TransonicFlow}, we can expect the flow in such a configuration changes from subsonic to supersonic when it passes through a surface with $z$ slightly shifted from the position of the smallest $S$ by the small effect of the gravitational force.

Since the specific energy, $\varepsilon_{\rm up, in}$, of the matter in the upper layer should be kept even after the flow-direction-change, the Bernoulli equation for the flow along the $z$ axis is expressed as
\begin{equation}
\frac{v^{2}}{2} + \frac{1}{\gamma-1} c_{\rm s}^{2} -\frac{GM}{z} \simeq \varepsilon_{\rm up, in}
\label{eqn:BernoulliEq-NR}
\end{equation}
in the non-relativistic case.
If the flow gets sufficiently supersonic, the terminal velocity of the flow, $v_{\infty}$, is given as
\begin{equation}
v_{\infty} \simeq \sqrt{2 \varepsilon_{\rm up, in}}.
\label{eqn:TerminalV}
\end{equation}
When $\varepsilon_{\rm up, in} \gtrsim c^{2}$, we have to take account of the relativistic effect and see the terminal Lorentz factor of the flow, $\Gamma$, which is approximated as
\begin{equation}
\Gamma \simeq 1 + \frac{\varepsilon_{\rm up, in}}{c^{2}}.
\label{eqn:TerminalGamma}
\end{equation}
Since $\varepsilon_{\rm up, in}$ can be as large as $\sim c^{2}$ or even larger, the flow is expected to turn to the relativistic jets with $\Gamma \gtrsim 2$.

The energy flow from the bottom layer of the slim disk to the final relativistic jets in the present scenario is summarized as follows:
\begin{enumerate}
\item[1)] In the radiation-pressure dominated situation of the slim disk, radiative energy flow via the diffusion process along the pressure gradient arises in the vertical direction, and the energy flux from the bottom layer is almost all absorbed by the upper layer than a critical height.
\item[2)] The energy absorbed in the upper layer is integrated in association with the matter in-fall and the accumulated specific energy can be as large as $\sim c^{2}$ in the vicinity of the black hole when the accretion rate through the upper layer is around $\dot{M}_{\rm c}$.
\item[3)] A large fraction of the matter in the upper layer turns the flow direction to outward along the central axis of the accretion disk, keeping the specific energy $\sim c^{2}$.
\item[4)] The outward flow becomes sufficiently supersonic and almost all the specific energy is converted to the kinetic energy at infinity, making the terminal speed relativistic.
\end{enumerate}


\section{Observational indications to accretion flows for transient jet ejections}
We have seen above that a slim disk could be possible to provide a situation in which the upper layer gets positive specific energy as large as $\sim c^{2}$ and finally turns to relativistic jets.
Now, let us see whether observations fit the scenario in which the slim disk plays the key role for the transient, relativistic jet-ejections.

Mirabel et al. (1998) performed simultaneous observations of X-rays, infrared, and radio wavelengths of GRS 1915+105 and detected the ejection of relativistic
plasma clouds in the form of synchrotron flares at infrared and radio wavelengths. From the observational results, they infer that the onset of the ejection takes place at the time of a spike appearing when the state changes from the hard state  to the soft state.

According to the definitions by Belloni et al. (2000), these hard state and soft state can be regarded as the state C and the state A followed by the state B respectively, 
and the intensity spike is seen just on an occasion of the transition from the state C to state A.

Belloni et al. (2000) systematically investigated time-variabilities of GRS 1915+105 and classified the patterns into 12 classes.
They further showed that the variability-patterns can be understood to be composed of transitions between three basic states, A, B and C.
An interesting finding by them is that GRS 1915+105 in 1990's
moves between the three states, A, B and C following certain number of patterns.
A transition from C to B never occurs.  
Thus, mutual transitions between A and B and between A and C happen 
but those between B and C never appear.  
When C changes to B, it is always done on a way from C via A to B.  
A cycle with the order of C - A - B is often repeatedly seen in various patterns.

The ejection timings of the transient jets from GRS 1915+105 look always consistent with the moment of the C to A transition in the C - A - B cycles (Mirabel et al. 1998; Klein-Wolt et al. 2002).
Presence of continuous short period (20 -- 40 min) oscillations in the radio flux from GRS 1915+105 was reported by Fender et al. (1999), while the repeated C - A - B cycles on such timescales are often observed in the X-ray observations (Belloni et al. 2000).
This could be consistent with the picture in which the mass ejection always appears in every C to A transition more or less.
Thus, we could say that the transient jet ejection happens at the moment of the C to A transition in the C - A - B cycle.

Investigating the flux- and spectral changes in the C - A - B cycle of GRS 1915+105, Inoue (2022a) discusses the configuration-variations of the accretion flow in the C - A - B cycle. 
The discussions are reviewed in Appendix \ref{ObsIndications}, where one unit of the C - A - B cycle is divided into four phases, the phase I to IV, and the C to A transition is shown to appear in association with the transition from the phase III to IV.

In the first place, each of the C - A - B cycle is conjectured to be basically governed by a limit cycle between the gas-pressure dominated standard disk (GDSD) and the slim disk, which appears in a region with a fairly large distance from the central black hole. 
In this conjecture, a unit cycle starts from the transition from the GDSD to the slim disk at a position in the limit-cycle-triggering region and the slim disk with relatively high accretion rate rapidly in-falls onto the black hole in the Phase I.
Then, the accretion rate through the limit-cycle-triggering region becomes very low and gradually returns to the initial value in the phase II through IV.
In this latter period, the replenishing front of the mass flow is considered to gradually advance inward from far outside of the limit-cycle-triggering region.

\begin{figure}
 \begin{center} 
  \includegraphics[width=14cm]{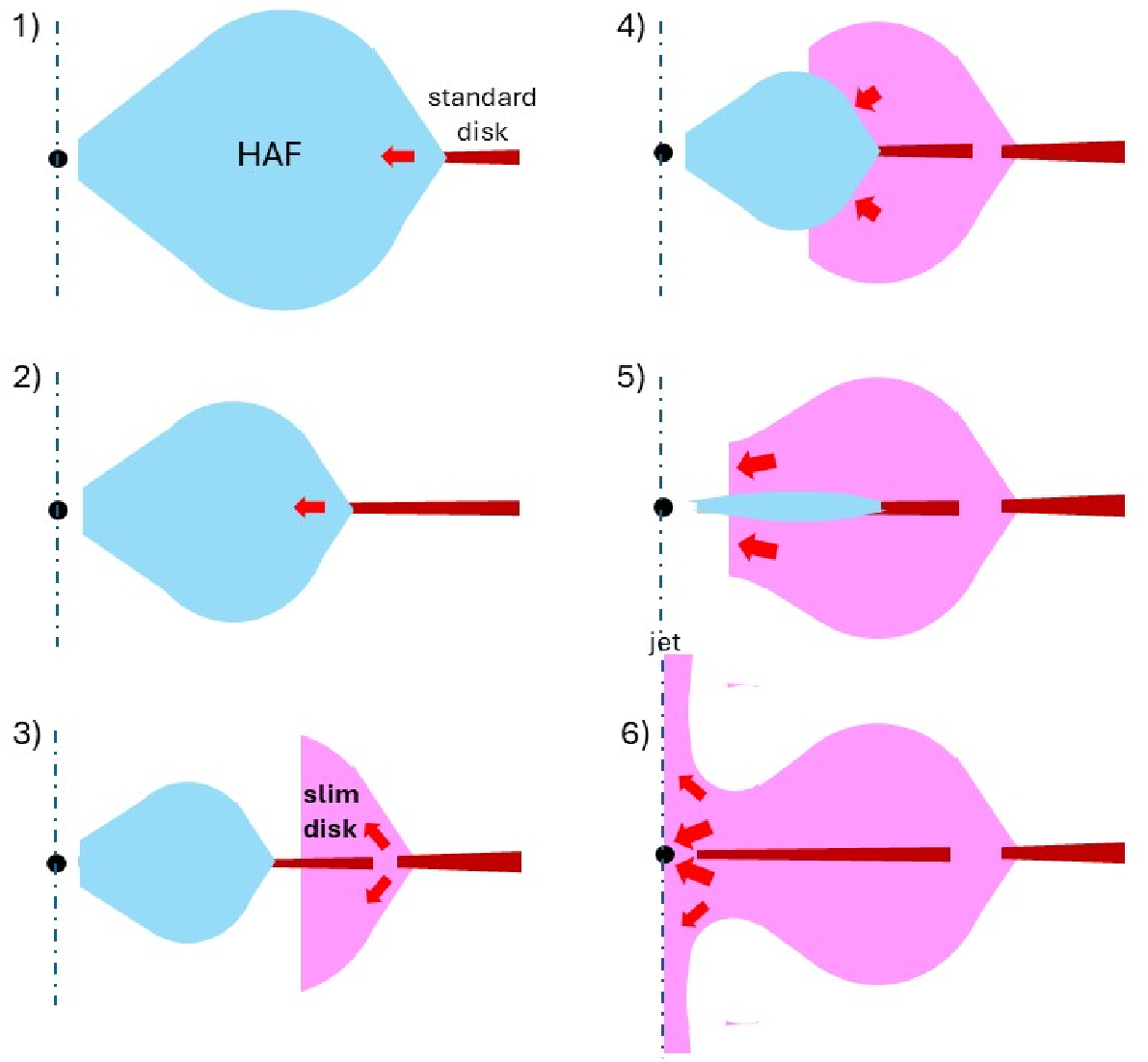} 
 \end{center}
\caption{Sequence of the flow-configuration-change in which a slim disk emerges from a standard disk and its upper layer turns to a jet.  For explanations of the panel 1 through 6, see the text.}\label{fig:FlowEvolution}
\end{figure}

The limit cycle is triggered when the accretion rate through a GDSD at a certain position exceeds the boundary rate between the GDSD and the radiation pressure dominated standard disk (RDSD) which is unstable.
The boundary position inside of which the GDSD changes to the RDSD is given to be proportional to $\dot{M}^{16/21}$ in Equation (\ref{eqn:r_b}).
Thus, another limit cycle is possible to arise in a region much closer than the initial limit cycle region even if the accretion rate is on the way of the recovery.
The observational evidences in the phase III to IV as discussed in Appendix \ref{ObsIndications} could indicate that another limit cycle really appears with a smaller scale on an inner side than the initial limit cycle, providing an occasion for the slim disk to play the key role for the jet ejection.
The sequence of the flow-configuration-change, in which the mechanism of the relativistic jet ejection from the slim disk presented in Section \ref{Scenario} is realized, is described as follows.
Figure \ref{fig:FlowEvolution} exhibits schematic views of the flow-configuration-change for the reference. 

\begin{enumerate}
\item[1)] The replenishing front of the mass flow from the companion star advances inward in Phase III.  The innermost part of the front is the HAF with very low accretion rate and the standard disk with relatively high accretion rate surrounds it.
\item[2)] The boundary between the HAF and the standard disk moves inward as the accretion rate through the standard disk increases.
\item[3)] The accretion rate through the standard disk at a certain position enters in the unstable range and the standard disk turns to the slim disk there.
\item[4)] The slim disk infalls toward the black hole and contacts with the HAF.
\item[5)] The slim disk crushes the HAF and reaches to the vicinity of the black hole.
\item[6)] The upper layer of the slim disk turns to the jet as discussed in the previous section.
\end{enumerate}

\section{Discussion}\label{Discussion}
We have discussed that it can be interpreted how the jet matter gets positive specific energy as large as $\sim c^{2}$ by considering the  energy transfer through radiative diffusion from the bottom layer to the upper layer in a slim disk.
In this scenario, the accretion rate through the upper layer is required to be $\sim \dot{M}_{\rm c}$, which is about $2 \times 10^{18}$ g s$^{-1}$ for $M = 10 M_{\odot}$, for the matter in the upper layer to have the specific energy of $\sim c^{2}$ and to form the relativistic jets.
The accretion rate through the bottom layer in this situation is about 40 $\dot{M}_{\rm c}$ as seen in Figure \ref{fig:Mdot-Pb} and thus we can see that only 1/40 of the specific energy of the matter in the bottom layer is devoted to the upper layer, causing the relativistic jet ejection.

If almost all the matter in the upper layer turns to the jet flow, the accretion rate through the upper layer can be considered to be the jet mass-flow rate.
As discussed in Appendix \ref{JetEjection}, the jet mass-flow rate is roughly evaluated a few times 10$^{18}$ g s$^{-1}$, which is consistent with the above value required in the present scenario.

The evaluation of the jet mass-flow rate in Appendix \ref{JetEjection} is done based on the estimate for the jet duration to be a few sec.
This duration can be considered to be the time in which the slim disk comes into existence at some place with a distance, $r_{\rm out}$, and falls on to the black hole.
The time, $t_{\rm slm}$, could roughly be estimated to be $\sim r_{\rm out}/v$.  
If we approximate the in-falling velocity as $v \sim \alpha c_{\rm s}^{2}/v_{\rm k}$, where $\alpha$ and $v_{\rm k}$ are the viscosity parameter and the Keplerian circular velocity at $r_{\rm out}$ according to the standard accretion disk theory (Shakura \& Sunyaev 1973) and assuming $c_{\rm s}^{2} \sim GM/(6r_{\rm out})$, we get 
\begin{equation}
t_{\rm slm} \simeq (6\sqrt(2)/\alpha) (r_{\rm out}/R_{\rm s})^{3/2} (R_{\rm s}/c),
\label{eqn:t_slm}
\end{equation}
which is a few sec for $r_{\rm out} \sim $several 10 $R_{\rm s}$ and $\alpha \sim 0.1$.

Since the duration of a few seconds is short observationally, jets appearing in such a short period are called transient jets.
This duration is, however, much longer than the dynamical time scale around the black hole and thus the transient jets can be regarded as a steady phenomena from a slim disk with a short life.
In fact, the present study shows that every slim disk can exhibit jet ejections unless the accretion rate through the bottom layer is as high as  $10^{3} \dot{M}_{\rm c}$ or higher.
It should be noted here that the steady flow configuration as discussed here is implicitly assumed to be stable but that it has not been confirmed yet. Future works on this point should be necessary. 

When the accretion rate is so high, the amount of the specific energy transfered to the upper layer is negligibly smaller than the intrinsic specific energy, as seen from Equation (\ref{eqn:varepsilon-up}) and nothing is expected to happen practically.
This situation can easily be understood as follows.

The energy flow in the slim disk is basically governed by Equation (\ref{eqn:EnergyFlowEq}).
The left hand side of this equation represents the energy flow rate in the $-r$ direction, the typical time scale of which would be the matter in-falling time, $t_{\rm if}$, roughly given as
\begin{equation}
t_{\rm if} \sim \frac{r}{v}.
\label{eqn:infalling_time}
\end{equation}
On the other hand, the right hand side expresses the energy flow rate in the $+z$ direction, 
the typical time scale of which would be the photon diffusion time, $t_{\rm df}$, across a typical disk thickness, $h$, approximately expressed as
\begin{equation}
t_{\rm df} \sim \frac{3 \kappa_{\rm T} \rho h^{2}}{c}.
\label{eqn:diffusion_time}
\end{equation}
This time scale is derived by obtaining a diffusion equation on $U$ from equations (\ref{eqn:PhotonDiffEq}) and (\ref{eqn:dotU}) as
\begin{equation}
\dot{U} = \frac{\partial}{\partial z} \frac{c}{3\kappa_{\rm T} \rho} \frac{\partial U}{\partial z},
\label{eqn:DiffEq-U}
\end{equation}
and by approximating this equation as $U/t_{\rm df} \sim Uc/(3 \kappa_{\rm T}\rho h^{2})$.

When the matter in-falling is much faster than the photon diffusion, the advection becomes very effective and the radiative energy transfer in the $z$ direction can be negligible.
Then, from Equations (\ref{eqn:infalling_time}) and (\ref{eqn:diffusion_time}), the ratio of $t_{\rm if}/t_{\rm df}$ is expressed in terms of the accretion rate, approximated as $\dot{M} \sim 4\pi r h \rho v$, as
\begin{equation}
\frac{t_{\rm if}}{t_{\rm df}} \sim \frac{2}{3} \frac{r}{h} \frac{r}{R_{\rm s}} \frac{\dot{M}_{\rm E}}{\dot{M}},
\label{eqn:t_if/t_df}
\end{equation}
where $\dot{M}_{\rm E} \equiv 4\pi GM/(\kappa_{\rm T} c)$.
From this equation, we see that the condition of $t_{\rm if}/t_{\rm df} \ll 1$ for the radiative energy transfer from the bottom layer to the upper layer to be negligible is realized when $\dot{M} \gg 10 \dot{M}_{\rm E}$, considering $h \sim z_{\rm b}$ and $r \sim 10 R_{\rm s}$.

As discussed in Appendix \ref{PhaseI}, the front of the slim disk is considered to approach to the vicinity of the black hole at the moment of the transition from the state B to the state C, similarly to the case of the transition from the state C to the state A.
However, no radio enhancement is observed right after the B to C transition, differently from the case of the C to A transition (see e.g. figure 1 in Fender et al. 1999).
This difference could be understood to be due to the difference of the accretion rate through the slim disk.

As seen from the typical light curve of the C - A - B cycle when the transient jet ejections are observed  (see e.g. Fig.20b in Belloni et al. 2000), the intensity-decay time of the B to C transition is about 10 times as long as that of the C to A transition.
Inferring that this time scale could corresponds to $t_{\rm slm}$ in Equation (\ref{eqn:t_slm}), $r_{\rm out}$ of the B to C transition could be estimated to be about $10^{2/3}$ times as large as that of the C to A transition.
The outermost position of the slim disk with the distance $r_{\rm out}$ can be considered to be $r_{\rm b}$ given in Equation (\ref{eqn:r_b}) and to be proportional to $\dot{M}^{16/21}$.
Hence, the accretion rate through the slim disk in the B to C transition is evaluated to be approximately $10^{7/8}$ times as large as that in the C to A transition.
If we consider that the C to A transition corresponds to the case of $\bar{\varepsilon}_{\rm up,in} \sim c^{2}$, the accretion rate in this case is $\sim$ 40 $\dot{M}_{\rm c}$ as discussed in Subsection \ref{AccretionRateRelation} and thus that in the B to C transition is calculated to be $\sim 3 \times 10^{2} \dot{M}_{\rm c}$.
Then, we see from figure \ref{fig:Mdot-Pb} that $\bar{\varepsilon}_{\rm up, in}$ is $\sim 0.03 c^{2}$ for $\dot{M}_{\rm btm, in} \sim 3 \times 10^{2} M_{\rm c} $.
Since this $\bar{\varepsilon}_{\rm up, in}$ value is smaller than the gravitational energy in the vicinity of the black hole, no powerful mass ejection so as to exhibit noticeable observational evidences could be expected in this case.

When the accretion rate through the bottom layer is $\sim 10^{2} \dot{M}_{\rm c}$ as indicated with the upper horizontal dashed line in Figure \ref{fig:Mdot-Pb}, jets with the terminal velocity of 
$\sim 0.4\ c$ calculated from Equation (\ref{eqn:TerminalV}) for $\varepsilon_{\rm up, in} = 0.1$
and the mass ejection rate of 10 $\dot{M}_{\rm c}$ is expected.
This case seems to correspond to the case of jets observed from SS433.  From the observations, the jet velocity is known to be 0.26 $c$, the jet mass-ejection rate and the total accretion rate through the slim disk are inferred to be of the order of $10^{19}$ g s$^{-1}$ and of the order of $10^{20}$ g s$^{-1}$ respectively (see Inoue 2022b and references therein), which are roughly consistent with the values for $M \sim 10\ M_{\odot}$ in the case of the upper dashed line in Figure \ref{fig:Mdot-Pb}.
Although Inoue (2022c) already proposed the jet ejection mechanism for the jets of SS433 by considering the radiative energy transfer from the bottom layer to the upper layer in a slim disk, the present study has deepen the idea.


Next, we have discussed on how the front region of the slim disk which has been inflowing towards the central black hole on the equatorial plane turns to the jet flow along the $z$ axis normal to the disk plane.
Since the matter in the front region gets the specific energy much larger than the gravitational energy as discussed above, it is very reasonable that the matter starts expanding in the $z$ direction.
In parallel, it is conjectured that the angular momentum of the innermost matter is removed out and the matter concentrates around the $z$ axis well outside the black hole boundary.  
Then, the pressure around the central $z$ axis is expected to be enhanced so as to brake the inward motion in the $r$ direction.

The bounce of the flow in the $r$ direction around the $z$ axis in association with the movement toward plus $z$ direction makes the derivative of the cross section of the outward flow along the $z$ axis change from minus to plus as the matter moves, which causes for the flow to get supersonic as studied in Appendix \ref{TransonicFlow}.

The several properties of the transient jets from GRS 1915+105 have been shown above to be basically consistent with the present scenario.
A question could be whether this slim-disk model can be applied to transient jets from other black hole binaries in general.
The ejections of the discrete relativistic jets are also observed from several recurrent transient X-ray sources (Fender et al. 2004; 2009).

The transient jets from both from GRS 1915+105 and the recurrent transient X-ray sources appeared when the sources were in the very high state which is characterized by the various time variabilities on the time scale longer than 10 $\sim$ 10$^{2}$ s (see e.g. section 5 in Inoue 2022a).

The presence of the very high state was first pointed out from observations of GX 339-4 by Miyamoto et al. (1991).
They found that the X-ray intensity was at the highest level ever achieved and that the hardness - intensity diagram exhibited three localized domains: a hardness ratio increasing branch, a hardness ratio decreasing branch, and their crossing region.  The power density spectrum was shown to have three different types, corresponding to the three domains.
The correlated X-ray timing and spectral behaviors of GRS 1915+105 were investigated by Chen, Swank and Taam (1997) and the obtained hardness - intensity diagram and power density spectra looked very similar to those of GX 339-4 in the very high state.
The patterns of the time variabilities of GRS 1915+105 in the very high state were systematically investigated and be understood to be composed of transitions between the three basic states, A, B and C by Belloni et al. (2000).
Kubota and Makishima (2004) analyzed the data from the transient source XTE J1550-564 and discussed that the period around the outburst peak could be divided into three regimes, the standard regime, the anomalous regime and the apparently standard regime by the spectral behaviors.
Kubota and Done (2004) added a sub-state called as the strong very high state to the three regimes in the very high state.
Belloni et al. (2005) precisely investigated the hardness - intensity evolution and the properties of the fast time variability of GX 339-4 by analyzing a large number of the observations.
Based on the results, they discussed that the very high state should be divided into two states, the hard intermediate state and the soft intermediate state.
The anomalous state was further introduced on the highest intensity and the softest hardness side (Belloni 2010).

As seen above, the sources exhibit various patterns of time variation of X-ray intensity and spectrum in the very high state.
Although the classification names diverge depending on the researchers' points of view, these time variations have several common features:
The first is the very high X-ray intensity that is close to the highest level of the respective sources.
The second is the repeated transitions between two states.
The transitions sometimes take place between the high intensity state and the low intensity state.
They were observed from GX 339-4, called as ``flip-flop" or ``dips" (Miyamoto et al. 1991), and the similar phenomena were seen in some of the 12 classes of GRS 1915+105 introduced by Belloni et al. (2000).
The repeated transitions often happen between the soft spectral state and the hard spectral state, too.
They are observed near the outburst peak of the transient black hole binaries (see e.g. Fender et al. 2004;  2009), and also in the cyclic behaviors between the A, B and C states of GRS 1915+105 as discussed later.
The third is the transient jet ejections.
They are often observed near the outburst peak of several transient black hole binaries including GX 339-4 and XTE J1550-564 and are shown to appear in association with the transition from the hard state to the soft state (Fender et al. 2004; 2009).
The transient jets from GRS 1915+105 is also inferred to appear when the source changes from the hard state to the soft state.

Although observational correspondences of the transitions between the sources are not sufficiently clear yet, the commonalities could indicate that the transient jets observed from GRS 1915+105 and the transient black hole binaries take place under the common physical processes to all the sources.

The scenario for the transient jet ejections presented in this paper can basically interpret several properties of  the transient jets observed from the black hole binaries.
However, the considerations are still on several rough assumptions and approximations. Obviously, more detailed study is required.


\appendix 

\section{Observational indications to accretion flows for transient jet ejections}\label{ObsIndications}

\subsection{Flows in the C -- A -- B cycle}

A remarkable feature found in one of the patterns with the C - A - B cycles of GRS 1915+105 by Belloni et al. (1997b) is the correlation between the duration of the low-count-rate period (corresponding to the state C) and that of the following high-count-rate period (corresponding to the states A and B).  
No correlation is seen between the low-count-rate period and the previous high-count-rate period.
Belloni et al. (1997b) further discovered a strong correlation between the time-length of one low to high event and the largest innermost radius in the low-count-rate period obtained by spectral fitting with the disk-blackbody spectrum.
They inferred that the larger radius is realized when the larger mass in the inner part of the disk is lost and that the time needed for removal and refill gets longer.
If the amount of the removal mass can vary from event to event, no correlation between the low-count-rate time and the previous high-count-rate time can be understood.

Comparing the C - A - B cycle behaviors of the black hole source, GRS 1915+105 with the type II burst activities of the neutron star source, the Rapid Burster (MXB 1730-335), Inoue (section 5 in 2022a) discussed that the cyclic performances are both understood to be basically governed by the limit cycle behavior between the gas-pressure dominated standard disk and the slim disk. 

Here, we try to clarify how the configuration of the accretion flow on to the central black hole changes in association with the C - A - B cycle of GRS 1915+105 under the limit cycle scenario.

\begin{figure}
 \begin{center} 
  \includegraphics[width=14cm]{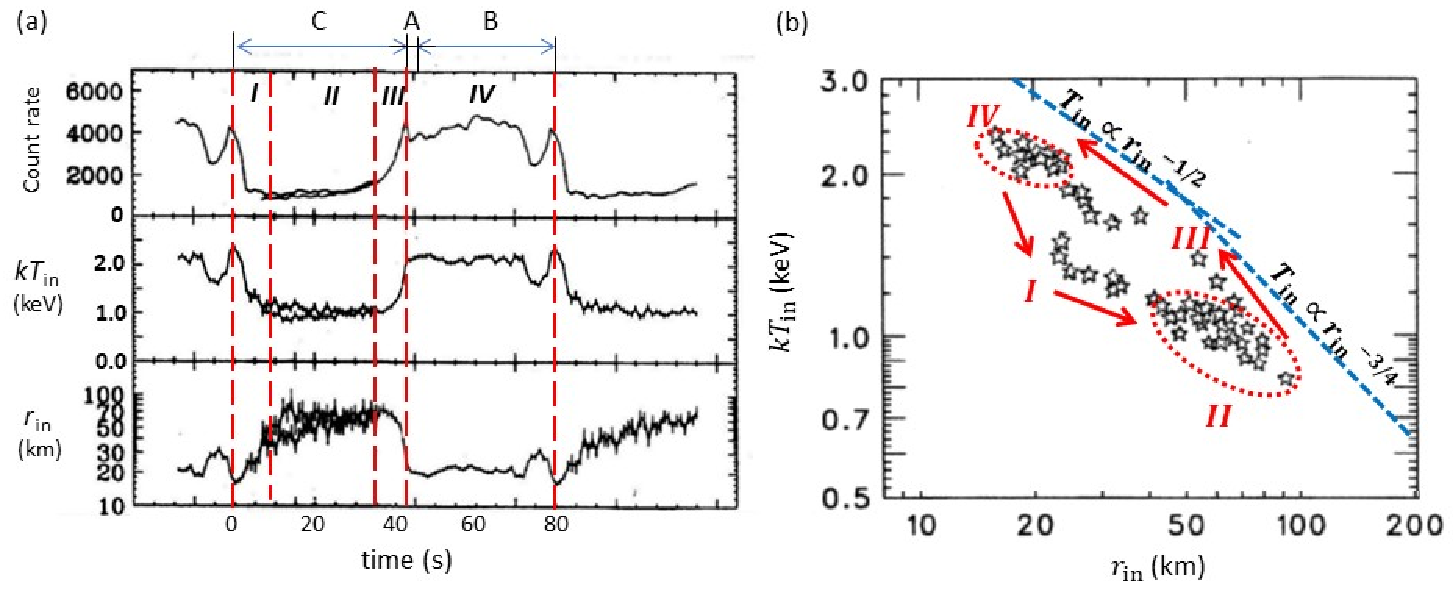} 
 \end{center}
\caption{(a) Typical X-ray light curve of the C - A - B cycle of GRS 1915+105  and the time variations of the innermost temperature and radius as the results of the model fits to the time-resolved spectra; (b) $r_{\rm in}$ -$T_{\rm in}$ diagram of the one cycle   [analyzed by Yamaoka (2001) and arranged by Inoue (2022a)]. }\label{fig:CyclicLC}
\end{figure}

Figure \ref{fig:CyclicLC}  shows the typical light curve of the C - A - B cycle approximately derived  from the original light curve obtained by Yamaoka (2001), to which the histories of the best fit values of the innermost temperature, $T_{\rm in}$, and radius, $r_{\rm in}$, are also  attached 
as the results of spectral fits to the 1-s resolved RXTE (the Rossi X-ray Timing Explorer) - data with the model composing of the disk-blackbody component and the broken power-law component.

It should be noted here that 
 $r_{\rm in}$ as the result of the spectral fitting of the model spectra to the observed spectra has a fairly large uncertainties.

One uncertainty is due to the dilution effect of the blackbody emission from the scattering dominated atmosphere (e.g. Shimura \& Takahara 1995), which makes the obtained $r_{\rm in}$ smaller than the real innermost radius.  

Another uncertainty could exist in the model fitting processes.
The relevant $r_{\rm in}$-change is obtained in the period C.
Since the spectrum in the period C is dominated by the power law component, the disk blackbody component, from which the $r_{\rm in}$ value is calculated, is obtained after the power-law component is subtracted from the observed spectrum.
However, we haven't exactly known the origin of the power law component and the exact spectrum on the low energy side.  
The uncertainty in the low-energy-side spectrum of the power law component could enhance the uncertainty of the disk blackbody component because of the dominance of the power law component to the disk blackbody component.
In particular, in case that $r_{\rm in}$ is as large as 100 km or so, the disk blackbody component is considered to exist in the energy range below 1 kev, which is the lowest end of the detector sensitivity.  Hence, it would be better to regard the derived temperature as the upper limit, and then the derived $r_{\rm in}$ becomes the lower limit.

These fairly large uncertainties in the estimation of $r_{\rm in}$ force us to see the estimated values of $r_{\rm in}$ as the rough indications of the innermost radius of the disk.

Inoue (2022a) divides the entire unit cycle of GRS 1915+105 into four phases, the phase I $\sim$ IV, in terms of the behavior of $r_{\rm in}$ variation.
As revealed in Figure \ref{fig:CyclicLC} - a, $r_{\rm in}$ increases from $\sim$20 km to $\sim$50-70 km in the phase I, stays around 50--70 km in the phase II, decreases from $\sim$50-70 km to $\sim$20 km in the phase III, and stays around 20 km in the phase IV.
According to the definition by Belloni et al. (2000), the phases I, II and III correspond to the state C, and the phase IV to states A and B.  The state A appears for a short period just after the start of the phase IV.

Here, let us draw phenomenological pictures, mainly following the discussions in sub-subsection 5.5.2 of Inoue (2022a), on what happens in the accretion flow in each of the four phases, looking carefully into movements of the data points on the $r_{\rm in}$ - $T_{\rm in}$ plane as exhibited in Figure \ref{fig:CyclicLC}-b.



\subsubsection{Phase I}\label{PhaseI}
A unit sequence of the C - A - B cycle is considered to start from an abrupt transition of the disk state from the standard thin disk to the slim disk at a certain position of the accretion disk. 
The matter in the region where the disk has changed to the slim disk would rapidly flow inward, running over the inner standard thin disk, and the accretion rate to the black hole should be enhanced then.

In the case of the neutron star source as the Rapid Burster, the enhanced accretion rate can be observed as the type II bursts as a result of the gravitational energy release at the stellar surface.
In the case of the black hole source as GRS 1915+105, however, the X-ray emission is observed not to be enhanced but to be rather suppressed then.
It could be due to absence of the solid surface and the situation of the slim disk, which is discussed in Section \ref{Discussion}.

It could be a natural conjecture that the phase I corresponds to the high $\dot{M}$ phase and the phase II $\sim$ IV appear in the low $\dot{M}$ phase in the limit cycle between the slim disk and the gas-pressure-dominated standard disk (see e.g. subsubsection 5.5.1 and figure 18 in Inoue 2022a).

The innermost radius of the accretion disk, $r_{\rm in}$, increases from $\sim$20 km to $\sim$50-70 km in the phase I.
This can be understood by considering the following situation.
The standard thin disk still exists on the inner side of the transition region soon after the disk transition.
The accretion rate through the thin disk, however, decreases since the matter of the standard disk in the transition region is evacuated by the rapid infall in the form of the slim disk.
If the accretion rate through the standard thin disk decreases, the inner region of a boundary radius turns to a HAF, and the boundary advances  outward associated with the accretion rate decrease.

The abrupt infall should, on the other hand, be accompanied by the rapid removal of the angular momentum from the infalling matter, and the surplus angular momentum should be carried by a matter outflow.
Hence, the accretion rate through the standard disk outer than the transition region is reduced by the matter outflow.

\subsubsection{Phase II}
In this phase, the flux is at the lowest level but slightly increases in time, while
$r_{\rm in}$ stays around 50 $\sim$ 70 km.
The region of $r\; <\; r_{\rm in}$ is considered to be occupied by a HAF because of the sufficiently low accretion rate
as in the low/hard state.
Indeed, the X-ray spectrum is dominated by the hard power law component in this phase (see e.g. Belloni et al. 1997a).
The outstanding power law component in the X-ray spectrum characterizes the low/hard state and is considered to arise via the Comptonization process from a two temperature HAF (e.g. Shapiro et al. 1976).
This phase could be explained as a period in which the steady flow to the transition region from the outside is interrupted by the transient matter outflow triggered by the disk transition. 

\subsubsection{Phase III}\label{PhaseIII}
$r_{\rm in}$ decreases while $T_{\rm in}$ increases in this phase. 
The situation of this phase can be considered in that the suppression of the accretion flow by the transient outflow carrying the dumped angular momentum has ceased and the replenishing front of the accretion flow is advancing inward.

The inner edge of the standard disk seems to move inwards but 
the negative logarithmic-slope of $T_{\rm in}$ against $r_{\rm in}$ seems to be different 
between in the former period and in the latter period of the phase III, as indicated with two different arrows in Figure \ref{fig:CyclicLC}-b.
The negative slope is first slightly larger than 3/4 and changes to $\sim$1/2.
As references, two typical negative slopes of 3/4 and 1/2 are exhibited with dotted lines in the figure.

The negative slope of 3/4 corresponds to a case when the accretion rate is constant through the standard disk.
The steeper negative slope than 3/4 in the former period suggests that the disk could be the standard disk and the accretion rate through the disk could gradually increase in this period.
The negative slope of 1/2, on the other hand, corresponds to a case when the luminosity is constant, since the luminosity from the optically thick disk is approximately proportional to $T_{\rm in}^{4} r_{\rm in}^{2}$.
The flattening of the data points in the latter period in the $r_{\rm in}$ - $T_{\rm in}$ diagram could suggest that the inner part of the standard disk turns to the slim disk in the mid of the phase III as the result of the $\dot{M}$ increase through the standard disk, and that the radiative loss gets to be suppressed by the dominance of the advection.

The appearance of the slim disk  is understandable by inferring such a transitional situation as follows.
The replenishing front of the accretion flow is advancing in this phase.
In the former period, the overall accretion rate distribution is well below the critical rate above which the gas-pressure diminated standard disk changes to the radiation-pressure dominated standard disk which is unstable.
A local accretion rate at some place in the standard disk, however, exceeds the critical rate as the result of the advance of the replenishing front, and the accretion flow turns to the slim disk there.

It should be noted here that the hard power law component is still dominant in the X-ray spectrum in this phase and is rather strengthened toward the end of the phase (see e.g. Figure 2 in Migliari \& Belloni 2003).
Since hot electrons in the HAF is considered to be responsible for the hard power law component, the significant presence of the hard power law component in the latter period of this phase suggests that the HAF coexists with the slim disk in this period.


The HAF could still exist in the innermost region of the standard disk remaining even in the latter half of the phase III on the inner side of the unstable region from which the slim disk has emerged.
In this case, the front of the slim disk is considered to run over the standard disk and to contact with the rising slope of the inner HAF at some distance, where the HAF is evolving from the initial thickness of the standard disk to the final one of the HAF.
Since the accretion rate through the slim disk is considered to be much larger than that through the HAF, the pressure of the slim disk would be much higher than that of the HAF initially and then the slim disk would press the rising slope of the HAF.
The compression would induce density-increase enhancing the radiative cooling and the temperature-decrease.  If we simply apply the approximations in the standard accretion disk theory as $h \propto \sqrt{T}$, $v \propto h^{2}$ and $\rho \propto h^{-1}v^{-1} \propto h^{-3}$ ($h$, $T$, $v$ and $\rho$ are respectively the height, the temperature, the inflow velocity and the density of the disk) to the HAF in the present situation, the positive feedback toward the denser and the slower flow is expected to be triggered, and eventually the expansion of the standard disk to the HAF could be quenched.
The intensity spike at the transitional moment from the phase III to the phase IV could be the result of the enhanced radiation associated with the compression of the HAF by the slim disk.

The present study is based on the inference that the transient jets are injected in the decay phase of the intensity spike.
If the HAF is really crushed by the slim disk, inducing the intensity enhancement, it would be impossible for the HAF to turn to the jets. 
Then, it would be reasonable to consider such a scenario for the slim disk to be responsible for the jet ejection as discussed in this paper.

\subsubsection{Phase IV}
The X-ray count rate increases toward the end of the phase III, and suddenly starts decreasing, which makes a spike in the X-ray light curve.
In parallel, $r_{\rm in}$ decreases toward the end of the phase III, and comes to  stay around 20 km, which is probably the innermost stable circular radius,  in the phase IV.

A short intensity dip appears after the spike, which corresponds to the state A characterized by the low intensity and the soft spectrum.
This state can be considered to be in a situation that the HAF and the slim disk have both disappeared and that only a standard thin disk extends down to the innermost stable radius.

After the dip, the count rate gradually returns to the level of the spike-peak, and the state B starts, in which the slim disk is inferred to exist in the inner region than the standard disk.

The data points in the phase IV in Figure \ref{fig:CyclicLC}-b distribute along the line with the negative slope of $\sim$1/2 extrapolated from the data points in the latter period of the phase III.
This strongly supports the picture that the slim disk is responsible for the disk blackbody emission in the state B.

A slim disk is considered to appear when a local accretion rate enters the unstable range of the standard disk at a certain position, 
and to end its life on the way of one limit cycle.
Hence, the state B having a fairly long duration could consists of a number of limit cycles on the very short time scales which happen here and there in the inner region of the accretion flow.
Indeed, the state B exhibits erratic intensity variations.

\subsection{Flows at the moment of jet-ejection}\label{JetEjection}
As repeatedly mentioned, the ejection of the transient jets is inferred to take place at the moment of a spike in the X-ray light curve between the periods of the state C and A.
According to the phenomenological arguments given above, the sharp intensity- increase towards the spike-peak could be due to the enhanced radiation associated with the crush of the HAF by the slim disk and the jet ejection could happen in the decay phase of the spike.
The declining slope of the intensity spike at the transition from the state C to the state A seen in the light curves in Belloni et al. (2000)  indicate that the duration of the decay phase is a few seconds.

The ejected mass in a pair of transient jets are estimated to be $\sim 10^{19}$ g (Mirabel et al. 1998; see also Fender et al. 1997; 1999), and then the mass flow rate through the jets could be a few times $10^{18}$ g s$^{-1}$ if the jets are ejected in the a few sec duration.


\section{Transonic flow}\label{TransonicFlow}
We consider a steady flow along the $z$ direction which is axially symmetric around the $z$ axis.
If the velocity-component perpendicular to the $z$ direction is relatively small, we can approximate it as a one-dimensional flow by introducing the cross section of the flow, $S$, as a function of $z$.
Then, we can get from the mass-flow-rate conservation as
\begin{equation}
\frac{1}{v_{\rm z}} \frac{dv_{\rm z}}{dz}+\frac{1}{\rho}\frac{d\rho}{dz}+\frac{1}{S}\frac{dS}{dz}=0,
\label{eqn:ConstFlowRate}
\end{equation}
where $v_{\rm z}$ and $\rho$ are the flow velocity in the $z$ direction and the density of the flowing matter.
The equation of motion is given as
\begin{equation}
v_{\rm z}\frac{dv_{\rm z}}{dz} = -\frac{1}{\rho}\frac{dP}{dz} - \frac{GM}{z^{2}},
\label{eqn:MotionEq}
\end{equation}
and the pressure gradient term can be rewritten in terms of the sound velocity, $c_{\rm s}$ as
\begin{equation}
\frac{1}{\rho}\frac{dP}{dz} =\frac{c_{\rm s}^{2}}{\rho} \frac{d\rho}{dz},
\label{eqn:dPdzTerm}
\end{equation}
using the polytropic relation between $P$ and $\rho$.
From the above three equations, the following equation is calculated as
\begin{equation}
(v_{\rm z}^{2} - c_{\rm s}^{2}) \frac{1}{v_{\rm z}}\frac{dv_{\rm z}}{dz} = \frac{c_{\rm s}^{2}}{S}\frac{dS}{dz} - \frac{GM}{z^{2}}.
\label{eqn:TransonicCond}
\end{equation}
We see from this equation that the transonic point is expected to appear at the point where the right hand side becomes zero.
The condition for the transonic point is given as
\begin{equation}
\frac{d\ln S}{d\ln z} = \frac{GM/z}{c_{\rm s}^{2}}.
\label{eqn:dSdzCond}
\end{equation} 
Considering the situation in which $c_{\rm s}^{2} \gg GM/z$, the sonic point is located at a position where the logarithmic gradient of $S$ against $z$ is slightly larger than zero.

In the present situation, the time derivative of $S$ within a radius $r$ in the comoving frame with the flow is expressed as $dS/dt = 2\pi r dr/dt$, which can be rewritten as
\begin{equation}
\frac{dS}{dz} = 2\pi r \frac{v_{\rm r}}{v_{\rm z}}
\label{eqn:dSdz}
\end{equation}
by using equations of $dt = dz/v_{\rm z}$ and $dr/dt = v_{\rm r}$.
As discussed in Subsection \ref{FlowConversion}, the motion in the $-r$ direction of the upper layer of the slim disk is considered to bounce back to the $+r$ direction around the central $z$ axis in association with the expanding motion to $+z$ direction, in the flow considered here.
This means that $v_{\rm r}$ changes the sign from minus to plus so that $dS/dz$ simultaneously changes the sign from minus to plus as the flow advances outward along the polar axis.
Thus, we can expect that there exists a plane of the cross section minimum and that the flow turns to be supersonic near there.


\end{document}